\documentclass[aps,prc,twocolumn,nofootinbib,amsmath,showpacs,superscriptaddress,floatfix]{revtex4-1}
\usepackage{graphicx,epsfig,epstopdf,longtable}
\usepackage{mathptmx}
\usepackage{xcolor}
\usepackage[light,first]{draftcopy} % ,bottomafter
\usepackage[pdftex]{hyperref}
\usepackage[mathlines]{lineno}% Enable numbering of text and display math
\usepackage{wasysym}
\begin{document}
\title{Experimental $\gamma$-decay strength in $^{59, 60}$Ni compared with microscopic calculations}

\author{T.~Renstr{\o}m}
\email{therese.renstrom@fys.uio.no}
\affiliation{Department of Physics, University of Oslo, N-0316 Oslo, Norway}
\author{G.~M.~Tveten}
\affiliation{Department of Physics, University of Oslo, N-0316 Oslo, Norway}
\author{J.~E.~Midtb{\o}}
\affiliation{Department of Physics, University of Oslo, N-0316 Oslo, Norway}
\author{H.~Utsunomiya}
\affiliation{Department of Physics, Konan University, Okamoto 8-9-1, Higashinada, Kobe 658-8501, Japan}
\author{O.~Achakovskiy}
\affiliation{Institute for Physics and Power Engineering, 249033 Obnisk, Russia}
\author{S.~Kamerdzhiev}
\affiliation{National Research Centre "Kurchatov Institute", Moscow, Russia}
\author{B.~Alex Brown}
\affiliation{National Superconducting Cyclotron Laboratory and Department of Physics and Astronomy,
Michigan State University, East Lansing, Michigan 48824-1321, USA}
\author{A.~Avdeenkov}
\affiliation{Institute for Physics and Power Engineering, 249033 Obnisk, Russia}
\affiliation{DST Hydrogen Infrastructure Center of Competence (HySA Infrastructure), North-West University, Faculty of
             Engineering, Private Bag X6001, Potchefstroom 2520, South Africa}
\author{T.~Ari-izumi}
\affiliation{Department of Physics, Konan University, Okamoto 8-9-1, Higashinada, Kobe 658-8501, Japan}
\author{A.~ G\"{o}rgen}
\affiliation{Department of Physics, University of Oslo, N-0316 Oslo, Norway}
\author{S.~M.~Grimes}
\affiliation{Department of Physics and Astronomy, Ohio University, Athens, Ohio 45701, USA}
\author{M.~Guttormsen}
\affiliation{Department of Physics, University of Oslo, N-0316 Oslo, Norway}
\author{T.~W.~Hagen}
\affiliation{Department of Physics, University of Oslo, N-0316 Oslo, Norway}
\author{V.~W.~Ingeberg}
\affiliation{Department of Physics, University of Oslo, N-0316 Oslo, Norway}
\author{S.~Katayama}
\affiliation{Department of Physics, Konan University, Okamoto 8-9-1, Higashinada, Kobe 658-8501, Japan}
\author{B.~V.~Kheswa}
\affiliation{Department of Physics, University of Oslo, N-0316 Oslo, Norway}
\affiliation{University of Johannesburg, Department of Applied physics and Engineering mathematics, Doornfontein, 2028, South Africa}
\author{A.~C.~Larsen}
\affiliation{Department of Physics, University of Oslo, N-0316 Oslo, Norway}
\author{Y.-W.~Lui}
\affiliation{Cyclotron Institute, Texas A\&M University, College Station, Texas 77843, USA}
\author{H~.-T.~Nyhus}
\affiliation{Department of Physics, University of Oslo, N-0316 Oslo, Norway}
\author{S.~Siem}
\affiliation{Department of Physics, University of Oslo, N-0316 Oslo, Norway}
\author{D.~Symochko}
\affiliation{Institute of Nuclear Physics (IKP), Darmstadt University of Technology, 64289 Darmstadt, Germany}
\author{D.~Takenaka}
\affiliation{Department of Physics, Konan University, Okamoto 8-9-1, Higashinada, Kobe 658-8501, Japan}
\author{A.~V.~Voinov}
\affiliation{Department of Physics and Astronomy, Ohio University, Athens, Ohio 45701, USA}

\date{\today}

\begin{abstract}
\vskip 0.5cm

Nuclear level densities and $\gamma$-ray strength functions have been extracted for $^{59, 60}\rm{Ni}$, using the Oslo method on data sets from the $^{60}$Ni($^{3}$He,$^{3}$He$^{\prime}\gamma$)$^{60}$Ni and $^{60}$Ni($^{3}$He,$\alpha\gamma$)$^{59}$Ni reactions. Above the neutron separation energy, S$_n$, we have measured the $\gamma$-ray strength functions for $^{61}$Ni and $^{60}$Ni in photoneutron experiments. The low-energy part of the $^{59,60}$Ni $\gamma$-ray strength functions show an increase for decreasing $\gamma$ energies. The experimental $\gamma$-ray strength functions are compared with $M1$ $\gamma$-ray strength functions calculated within the shell model. The $E1$ $\gamma$-ray strength function of $^{60}$Ni has been calculated using the QTBA framework. The QTBA calculations describe the data above $E_{\gamma}\approx$ 7 MeV, while the shell-model calculations agree qualitatively with the low energy part of the $\gamma$-ray strength function. Hence, we give a plausible explanation of the observed shape of the $\gamma$-decay strength.
\end{abstract}

\maketitle

%-----------------------------------------INTRODUCTION----------------------------------------------%

\section{Introduction}
\label{sec:intro}

The $\gamma$-decay channel is ubiquitous in nuclear reactions. Its properties provides information on the nuclear structure and are vital for cross-section calculations for a broad range of applications. Nuclear level densities (NLDs) and $\gamma$-ray strength functions ($\gamma$SFs) are indispensable quantities in the description of the $\gamma$ decay of excited states in the quasi-continuum region. 

During the past decade, an unexpected enhancement in the $\gamma$SF at low $\gamma$ energies ($E_{\gamma} \leq$ 3-4 MeV)\cite{First_upbend} has been revealed in a series of light to medium-mass nuclei, ranging from $^{27}$Si~\cite{Si_Magne} to $^{138}$La~\cite{La_Vincent} and $^{151,153}$Sm~\cite{Simon}. Currently, the issue of determining the electromagnetic character of the low-energy enhancement is experimentally unresolved, although Compton polarization measurements~\cite{Jones} show that it is likely to be dominated by M1 transitions. Further, it has been shown from angular distributions of $\gamma$ rays that the enhancement is dominated by dipole transitions~\cite{AC_dipole}. 
Theoretical models attribute the low-energy enhancement to transitions within the single-particle continuum producing $E1$ radiation~\cite{E1_upbend} or to a reorientation of the spins of high-$j$ neutron and proton orbits producing $M1$ transitions~\cite{M1_upbend}.  

For one isotope, $^{60}$Ni, there are experimental indications that the low energy enhancement is due to $M1$ transitions~\cite{AVoinov_TSC}. 
%Still, it is important to note that the fact that $^{60}$Ni exhibits only positive-parity states below $\approx$ 4.5 MeV could have significant consequences for the two-step cascade method used to obtain these results. As discussed in Ref.~\cite{AVoinov_TSC}, it means that for the secondary $\gamma$ ray, $M1$ transitions are strongly enhanced compared to $E1$ transitions.

Recent experimental results on $^{68}$Ni~\cite{Wieland_pygme, Rossi_pygme}, have revealed an enhancement in the $E1$ strength in the energy region around the neutron separation energy ($S_n$). Such a pygmy resonance is shown to be poorly reproduced in traditional Quasiparticle-Random-Phase-Approximation (QRPA) calculations of the $E1$ strength. However, the excess strength reported in~\cite{Wieland_pygme, Rossi_pygme} is reproduced in Ref.~\cite{E1_micro}, where $E1$ strengths of $^{58, 68, 72}$Ni are calculated using quasi-particle time blocking approximation (QTBA) calculations. 

Moving to more neutron-rich Ni isotopes, both a low-energy enhancement in the $\gamma$SF and a pygmy resonance have the potential to significantly increase their radiative neutron capture cross sections and may affect the r-process~\cite{Lar10, PygmyNucleo, Goriely_E1QRPA}. This is particularly important with regard to $^{76, 81}$Ni, as their radiative neutron capture cross section greatly influences the flow of the weak r-process, as shown in a sensitivity study performed in Ref.~\cite{RSurman}.

%Novel ($\gamma$,n)-data for $^{60,61}$Ni-isotopes are also presented. These data are not only interesting in their own terms, but also provide an opportunity to assure that the normalization data available from RIPL-3 provide realistic results (fra Gry). 

In this work, we report on the measurements of the $\gamma$SFs and NLDs in $^{59,60}$Ni below $S_n$. The spin distributions of $^{59,60}$Ni as a function of excitation energy have been extracted from shell-model calculations. The experimental NLDs are compared with existing data from particle evaporation measurements~\cite{AVoinov_evap1, AVoinov_evap2, AVoinov_evapEPJ}. We also report on measurements of the 
$\gamma$SF for $^{60,61}$Ni extracted from photoneutron experiments. Being the only stable, odd nickel isotope,$^{61}$Ni was chosen as a substitute to $^{59}$Ni and we assume that it constitutes a good approximation to the odd, unstable isotope $^{59}$Ni. Combining the charged particle measurements and the photoneutron measurements, we cover an energy range $E_{\gamma} \in [1, 20]$ MeV of the $\gamma$SF. Shell-model calculations of the $B(M1)$-values of $\gamma$-ray transitions in quasi-continuum have been performed and are compared to the experimental $\gamma$SFs of $^{59, 60}$Ni. Results of QTBA calculations of the $E1$ component of the $\gamma$SF of $^{60}$Ni are also compared to the experimental data.

%-----------------------------------------EXPERIMENTAL INFO/DATA ANALYSIS----------------------------------------------%

\section{Experiments and data analysis}
\label{sec:exp}
%-------------------------------------------------------------------------------------------------------------------------------------------%
\begin{figure}[tb]
 \begin{center}
 \includegraphics[clip,width=\columnwidth]{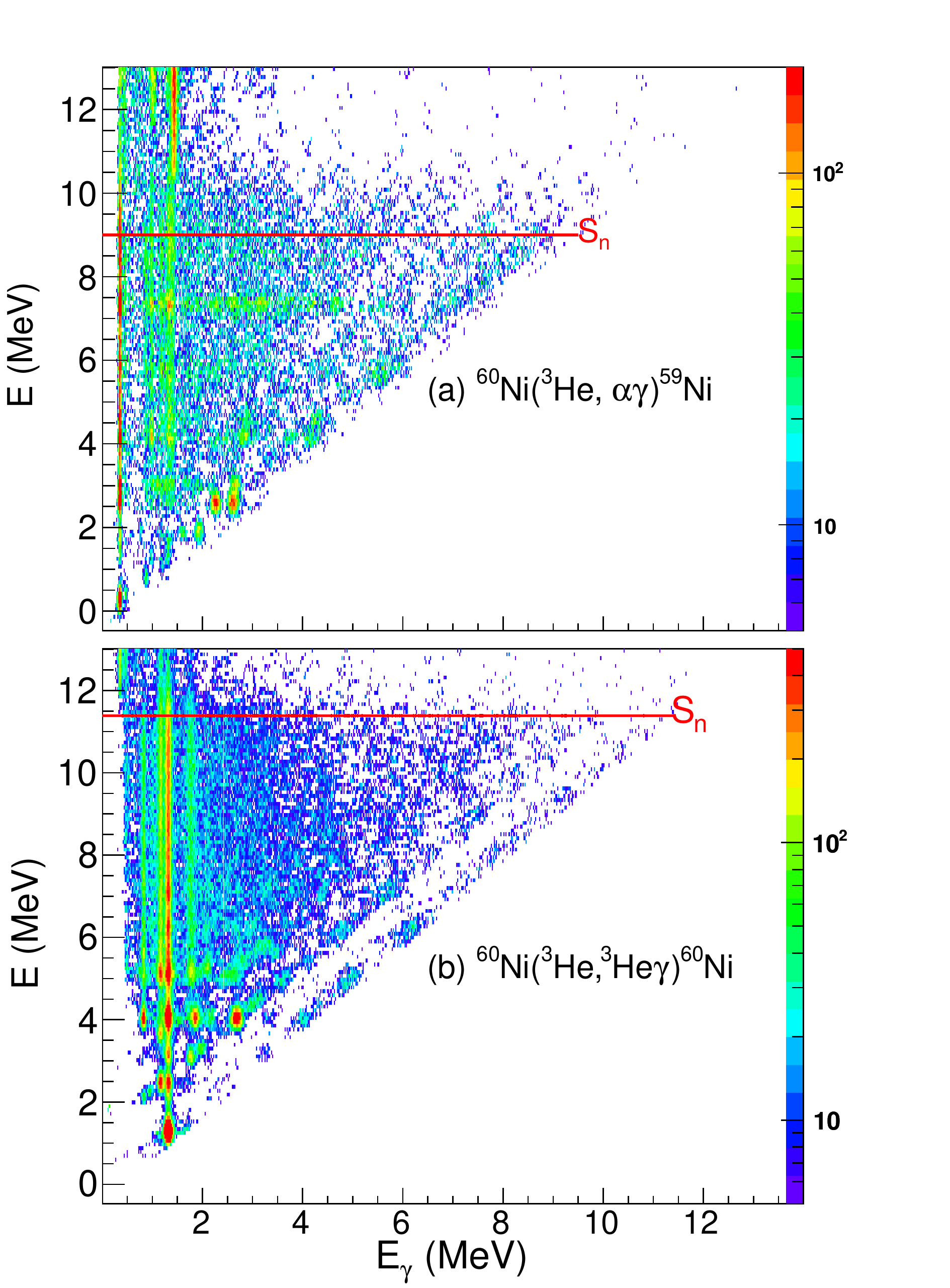}
%\vskip 2cm\includegraphics
 \caption{(Color online) Excitation energy E vs. $\gamma$ energy matrix for (a) $^{59}$Ni and (b) $^{60}$Ni. The $\gamma$-ray spectra are unfolded for each excitation-energy bin. The neutron separation energies, $S_n$, are indicated by horizontal lines.}
 \label{fig:unfoldplott}
 \end{center}
 \end{figure}
%-------------------------------------------------------------------------------------------------------------------------------------------%

The experiment was conducted at the Oslo Cyclotron Laboratory (OCL) with a $\approx$ 0.3 nA beam of 38~MeV $^{3}$He particles impinging on a 2.0 mg/cm$^{2}$ thick $^{60}$Ni, target enrichment (99.9$\%$) . Relevant particle-$\gamma$ coincidences were recorded in each of the analyzed reaction channels, namely $^{60}$Ni($^{3}$He, $^{3}$He$^{\prime}\gamma$) and the $^{60}$Ni($^{3}$He, $\alpha \gamma$). The charged ejectiles were identified and their energies measured with the silicon ring (SiRi) particle-detector system ~\cite{SiRi}. The SiRi detector consists of eight 130-$\mu$m silicon detectors, each divided into eight strips. One strip has an angular resolution of $\Delta\theta$ = 2$^{\circ}$. These segmented, thin detectors are placed in front of a 1550-$\mu$m thick back detector. In total, the SiRi system has 64 individual detectors, covering scattering angles between 40-54$^{\circ}$ and with a solid angle coverage of $\approx$6$\%$ of 4$\pi$. Using the known $Q$ values and reaction kinematics, the energy of the ejectile can be transformed into the initial excitation energy of the residual nuclei, in this case $^{59,60}$Ni.

The $\gamma$ rays were detected in 28 collimated 5$^{\prime \prime}$$\times$ 5$^{\prime \prime}$ NaI:Tl detectors, collectively called CACTUS~\cite{CACTUS}, surrounding the the target and particle detectors. The total efficiency of CACTUS is 15.2(1)$\%$ at E$_{\gamma}$ = 1332.5 keV. 

The collected data were sorted into total $\gamma$-ray spectra originating from different excitation bins. The resulting matrix constitutes the starting point of the Oslo method. For each excitation energy bin, the corresponding $\gamma$ spectrum was unfolded using the Compton-subtraction method described in Ref.~\cite{unfold} and corrected for the efficiency of the NaI detectors. The unfolded $\gamma$ spectra of $^{59,60}$Ni as a function of excitation energy E are shown in Fig.~\ref{fig:unfoldplott}. In the case of $^{60}$Ni there are two clear diagonals which represent transitions to the ground state and the first excited state, whereas for $^{59}$Ni, only one broad diagonal is visible, as it consists of transitions both to the ground state and the first excited levels. These $\gamma$ rays must stem from primary transitions in the $\gamma$-cascades.

One of the main features of the Oslo method is the extraction of the energy distribution of all primary $\gamma$-rays originating from various excitation energies. This is accomplished using an iterative subtraction technique~\cite{forstegen}, where the primary $\gamma$-ray distribution for a given excitation-energy bin $E_j$ is determined by subtracting a weighted sum of the $\gamma$ spectra for all the underlying bins $E_{i<j}$. This technique has been thoroughly tested~\cite{ACL_syst} and has been found to be reliable and robust when the $\gamma$-decay routes from a given excitation-energy bin are the same, regardless of how the states in that bin were populated; in this case, either directly via the inelastic scattering reaction or the pick up reaction, or indirectly from $\gamma$ decay of higher lying states.

\begin{figure}
\begin{center}
\includegraphics[clip,width=\columnwidth]{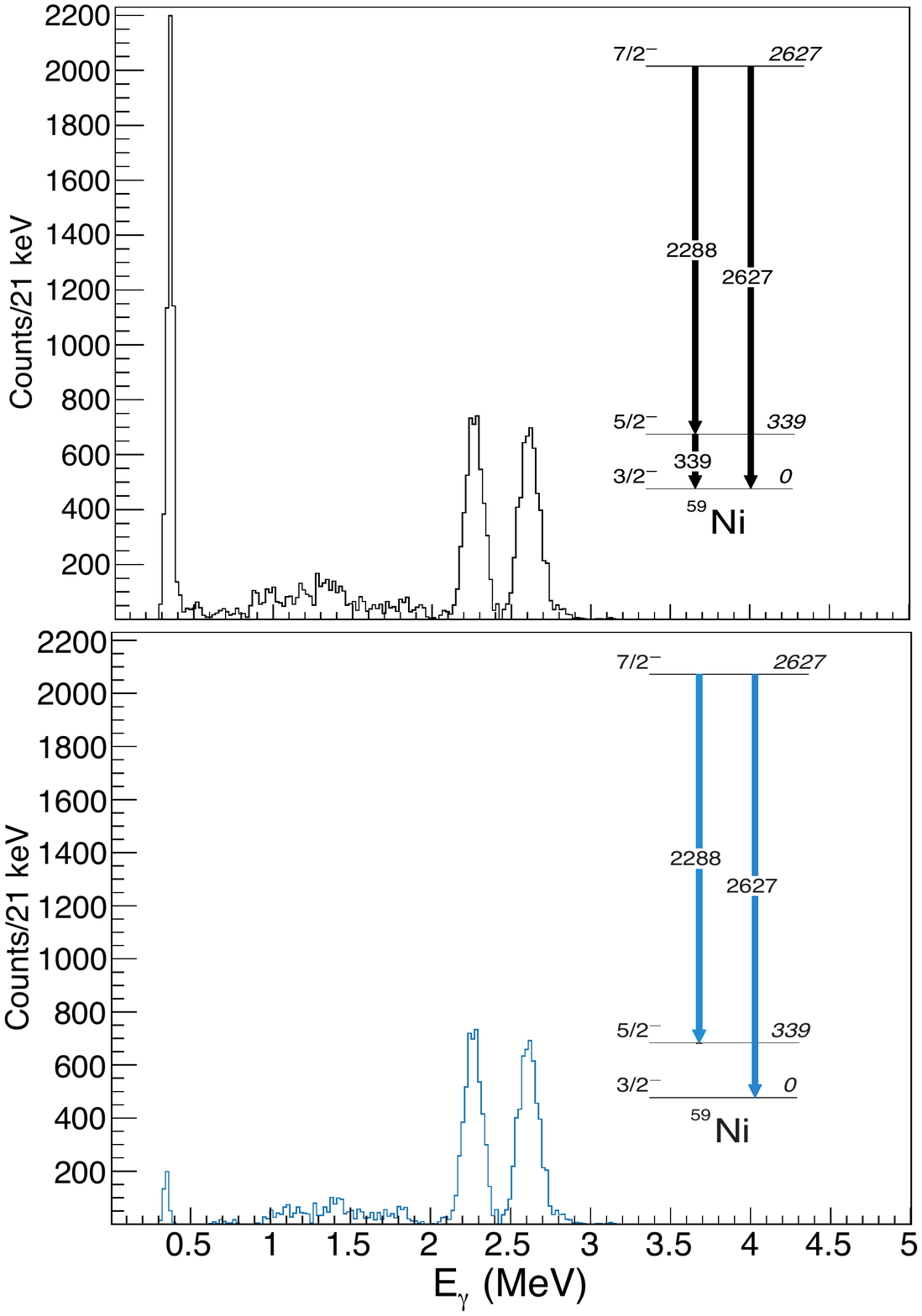}
\caption{(Color online) The upper panel shows the unfolded spectrum, while the lower panel displays the primary $\gamma$-ray spectrum.}
\label{fig:unf_first}
\end{center}
\end{figure}
 
In the following, we will illustrate how well we remove higher generation $\gamma$-rays from a certain excitation bin. For this purpose, we investigate the decay of a $7/2^{-}$ state at 2.63 MeV excitation energy in $^{59}$Ni. This state is populated with a rather high cross section in the ($^3$He, $\alpha$) reaction as seen in previous experiments as well~\cite{ni59_1,ni59_2,ni59_3}. Starting from this state, the nucleus will either decay directly to the ground state or to the first excited state and subsequently to the ground state. Our unfolded spectrum will contain all the three $\gamma$ rays from the two cascades as shown in the upper panel of Fig.~\ref{fig:unf_first}. The first generation $\gamma$ spectrum from the 2.63 MeV state consists of one 2.63 MeV and one 2.29 MeV $\gamma$-ray. After the first generation method has been applied, see Fig.~\ref{fig:unf_first} lower panel, the second generation $\gamma$-ray of 339 keV has been efficiently removed (only $\sim$ 8 \% remains). The intensities of the relative transitions are reported to be 100(19) and 75(10) respectively~\cite{NNDC}. In our case, the relative intensities of the 2.63 MeV and the 2.29 MeV peaks are measured to be $\sim$ 100 and 94 respectively, within the uncertainties of the listed values, indicating a good correction for the efficiency of the NaI detectors in the unfolding procedure.  
%We set a gate in the unfolded and the first generation matrix in the excitation energy range between $\approx$ 2.4 and 2.8 MeV (the FWHM of the first excited peak is about 300 keV in particle energy). The intensity of peaks is 100 and 94 respectively. This falls within the uncertainties of the listed branching intensities. 
% A 400 keV gate has ben applied to the 2.63 MeV state.

The primary $\gamma$-ray spectra represent the relative probability of a decay with $\gamma$-ray energy $E_{\gamma}$ from an initial excitation-energy bin $E_x$ and depend on the NLD at the final excitation energy $\rho(E - E_{\gamma})$ and the $\gamma$-ray transmission coefficient $\mathcal{T}(E_{\gamma})$ \cite{ASchiller_factor}:

\begin{equation}
P(E, E_{\gamma})  \propto \rho(E -E_{\gamma})  \mathcal{T}(E_{\gamma}),
\label{eq:factor}
\end{equation}
where $P(E, E_{\gamma})$ is the first generation matrix. The Oslo method has thus provided the functional form of the NLD and the $\gamma$SF. An iterative least $\chi^2$ method (described in Ref.~\cite{ASchiller_factor}) provides one solution to Eq.~(\ref{eq:factor}). Due to technical limitations of the method, we are not able to extract the $\gamma$SF for $\gamma$-ray energies below $\sim$ 1 MeV. In this work, we exclude $\gamma$-ray coincidences with energies below $\sim$ 2 MeV. Careful limits in excitation energy must also be applied in order to ensure the validity of Eq.~(\ref{eq:factor}). Here we have chosen to exclude coincidences with $E_{x} \leq 5$ MeV in the case of $^{59}$Ni and $E_{x} \leq 6.6$ MeV for $^{60}$Ni.

\section{Normalization of NLDs and $\gamma$SFs}
\label{sec:normalization}
After finding the functional form of $\rho$ and $\mathcal{T}$, we take into consideration that the products of the transformations 
\begin{eqnarray}
\tilde{\rho}(E-E_{\gamma})& = &A \textrm{exp}[\alpha(E-E_{\gamma})]\rho(E-E_{\gamma}), \\
\tilde{\mathcal{T}}(E_{\gamma})& = & B \textrm{exp}(\alpha E_{\gamma}) \mathcal{T}(E_\gamma),
\end{eqnarray}
also reproduce the first generation matrix. In the following, we will determine the transformation parameters A, B and $\alpha$ using existing neutron resonance data and level density information in the discrete energy range. This final part of the Oslo method is generally referred to as a $\it{normalization}$ of $\rho$ and $\mathcal{T}$.

\begin{figure}
\begin{center}
\includegraphics[clip, width = \columnwidth]{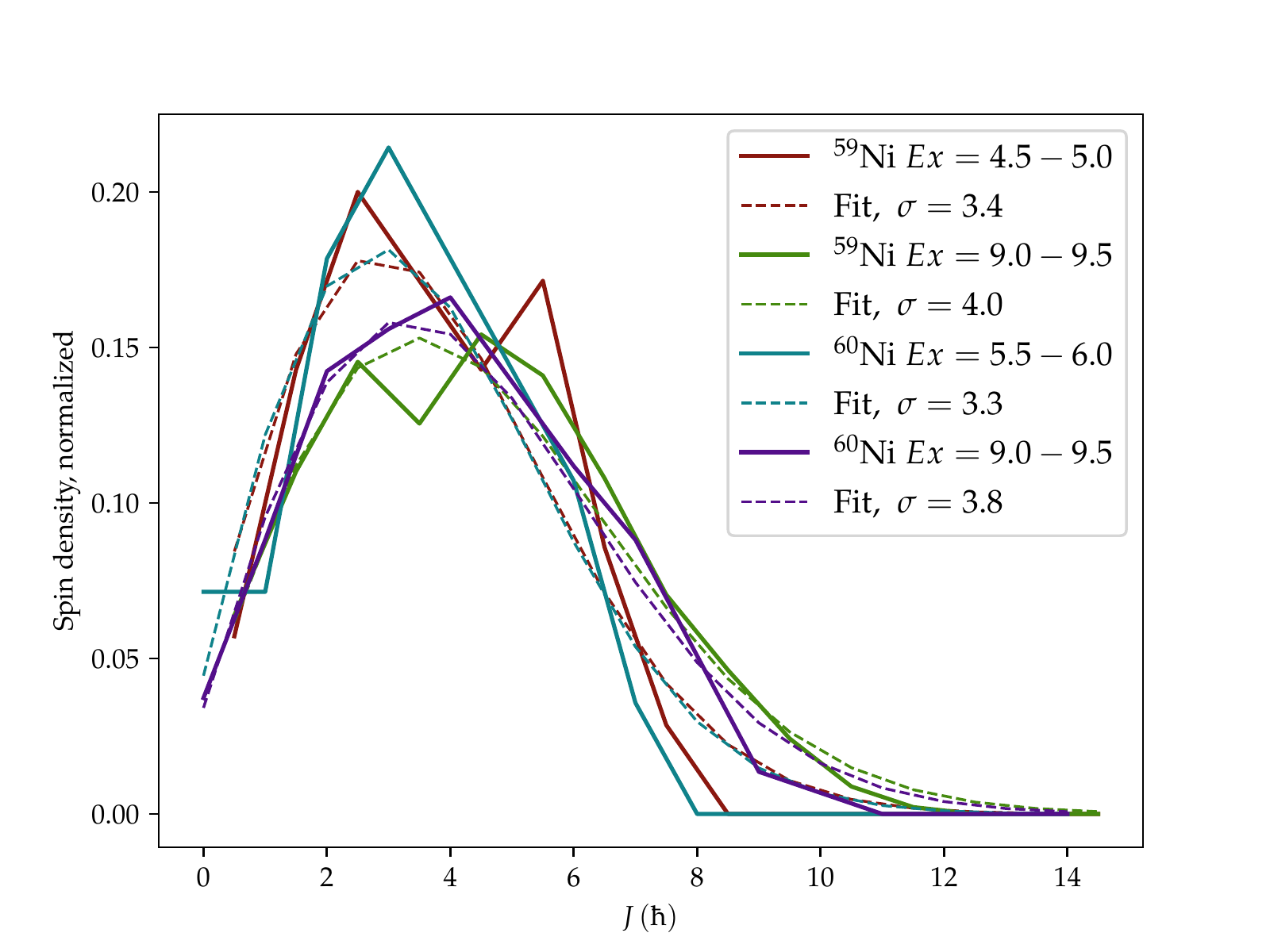}
%\vskip 2cm
\caption{(Color online) Examples of the spin distribution of calculated level densities for different excitation bins. Fits of the data with the spin distribution in Eq.(\ref{eq:spin_dist}) are also shown.}
\label{fig:spin_distribution_fit}
\end{center}
\end{figure}

\begin{figure}
\begin{center}
\includegraphics[clip, width = \columnwidth]{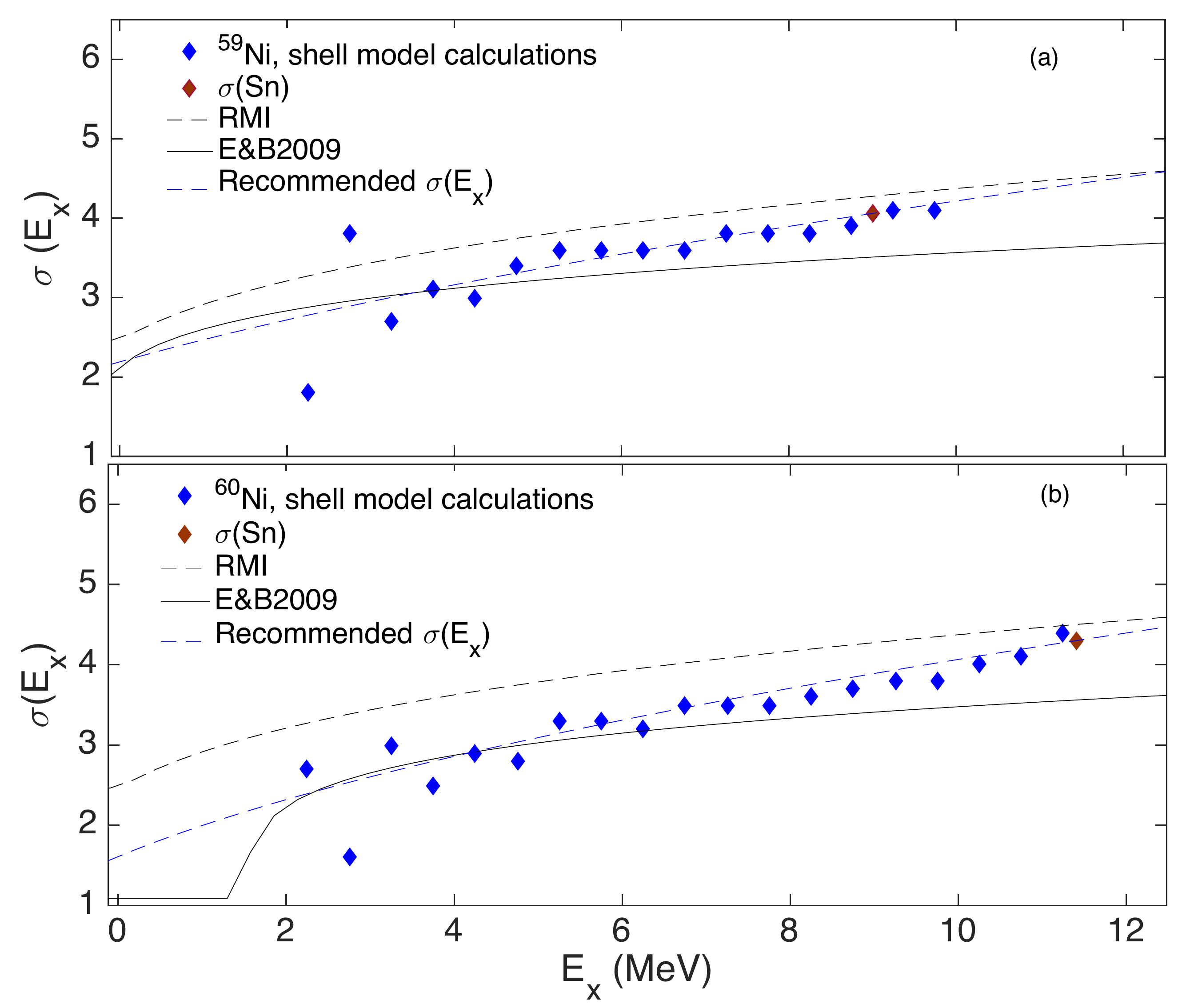}
%\vskip 2cm
\caption{(Color online) Spin cutoff values extracted from shell-model calculations (blue diamonds), together with values from the systematics reported in Ref.~\cite{E&B_2009} (full, black line) and in Ref.~\cite{E&B_2006} (dotted, blue lines). The red diamonds are the spin cutoff parameters used to estimate $\rho(S_n)$.}
\label{fig:spin_distribution}
\end{center}
\end{figure}

\subsection{NLDs}
\label{sec:level} 
\begin{figure}[tb]
\begin{center}
\includegraphics[clip, width=\columnwidth]{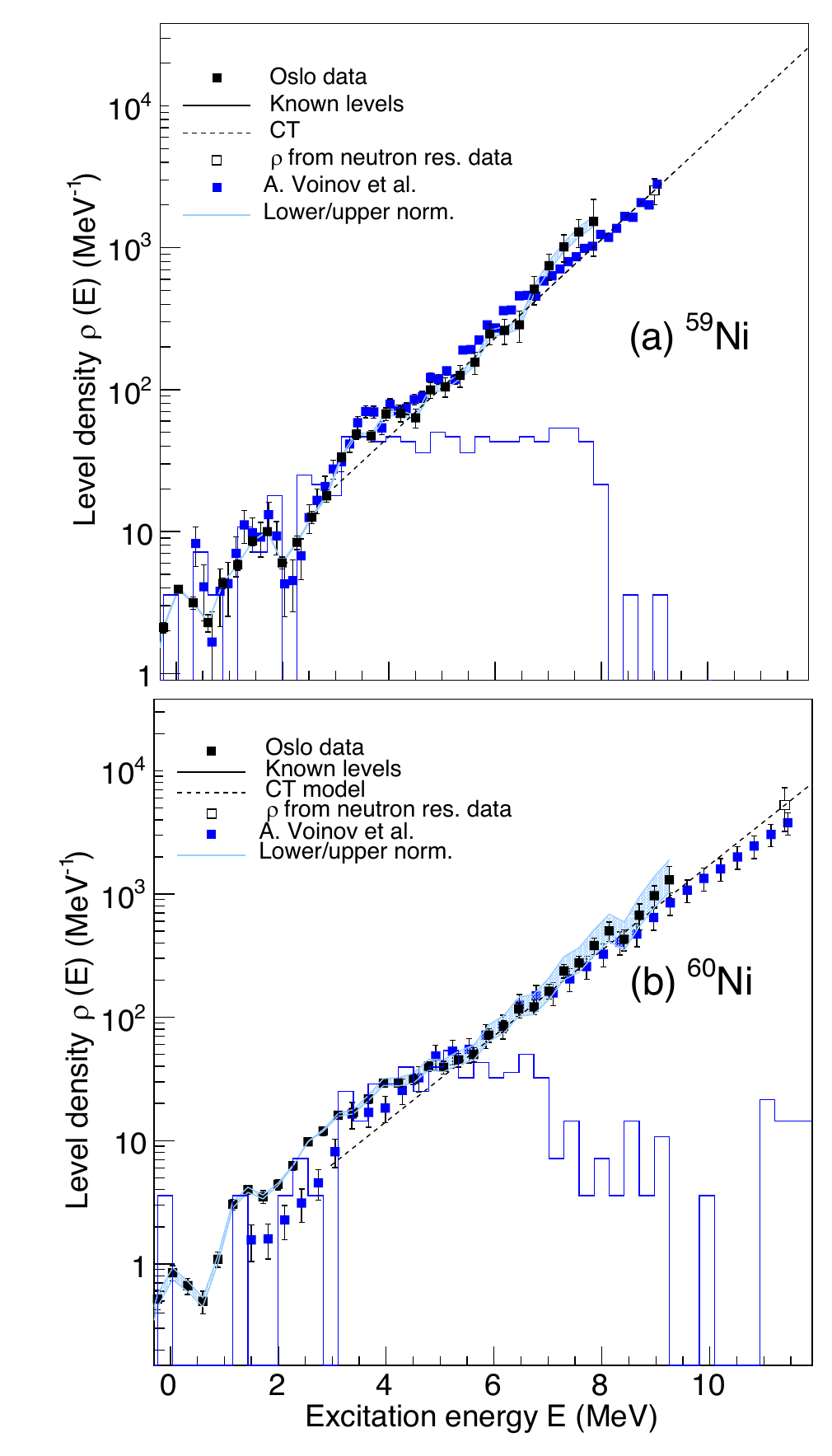}
%\vskip 2cm
\caption{(Color online) NLDs (black filled squares) of (a) $^{59}$Ni and (b) $^{60}$Ni, extracted using the Oslo method. The systematic uncertainties in the data are represented by light blue shaded areas. Our NLDs are compared with data from evaporation spectra ~\cite{AVoinov_evap1, AVoinov_evap2} (blue filled squares). The blue histograms represent the density of discrete states as listed in Ref.~\cite{NNDC}.}

\label{fig:oslo_voinov}
\end{center}
\end{figure}

For the normalization of the NLD we use information on the level density in the low-energy excitation range from discrete levels~\cite{NNDC}, see the blue histograms in Fig.~\ref{fig:oslo_voinov}(a) and (b), and the NLD at $S_n$. The level schemes of $^{59,60}$Ni are assumed to be practically complete up to excitation energies of 2.7 MeV and 4.6 MeV, respectively. 
In the excitation range around $S_n$, there exist average neutron resonance spacings~\cite{RIPL3}, constituting a partial level density: in the case of $^{58}$Ni(n, $\gamma$), capture of s-wave neutrons will lead to a population of $I^{\pi} = 1/2^{+}$ states in $^{59}$Ni. In the $^{59}$Ni(n, $\gamma$) reaction, $I^{\pi} = 1^{-}$ and $2^{-}$ states are populated in $^{60}$Ni. Thus, the average neutron resonance spacings provide the density of these specific populated spin/parity states right above $S_n$. Spin distributions at high excitation energies tend to be experimentally inaccessible, however in the case of $^{59,60}$Ni experimental data for certian exitation energies do exist, see Ref.~\cite{stevenGrimes, CCLu}. The estimation of the $\it{total}$ level density at $S_n$ from the partial ones has in our case been done using shell model calculations of the partial level densities of $^{59,60}$Ni. In this theoretical approach to studying the spin distribution of exited nuclei, we have used the \textsc{GXPF1A} $pf$ shell interaction~\cite{hamilton_1, hamilton_2}. The 300 first states of spin/parity $I^{\pi} \in \left [1/2^{-}, 29/2^{-}\right]$ and $I^{\pi} \in \left[0^{+}, 14^{+}\right]$ for $^{59,60}$Ni, respectively, have been calculated. The spin distribution was investigated by plotting the spin/parity dependent level densities as a function of spin for excitation energy bins of 500 keV. In the excitation energy region above $\sim$4 MeV the distributions are in good agreement with the phenomenological description of the spin distribution as proposed by Ericson ~\cite{TEricson}:
\begin{equation}
g(E,I) \simeq \frac{2I+1}{2\sigma^2}\mathrm{exp}[-(I+1/2)^2/2\sigma^2],
\label{eq:spin_dist}
\end{equation}
    with a single free parameter $\sigma$, usually referred to as the spin cutoff parameter. In Fig. \ref{fig:spin_distribution_fit} we show some extracted spin distributions from the shell-model calculations together with the fits of Eq.\ref{eq:spin_dist} to the data. We were able to estimate $\sigma$ from in an excitation energy range from $\sim$ 4 MeV up to $\sim$ $S_n$ of $^{59,60}$Ni, as shown in Fig.~\ref{fig:spin_distribution}. The estimated values of the spin cutoff parameter at neutron separation energy are $\sigma(S_n)$ = 4 and 4.3, for $^{59,60}$Ni, respectively. These values are in good agreement with the ones reported in~\cite{stevenGrimes, CCLu}.
 %----------------------------------------------------------------------------------------
\begin{table*}
\caption{Parameters used in normalization of the NLDs and transmission coefficients}
\centering
\begin{tabular}{l c c c c c c c c c c }
\hline
Nucleus & $I_t^{\pi}$     &    $S_n$    &  $\sigma(S_n)$  & $D_0$          &   $\rho(S_n)_R$   & $\langle \Gamma_{\gamma}\rangle$ \\  
 &                        &     (MeV)   &                 & (eV)           &    (MeV$^{-1}$)   &   (meV)                    \\ 
\hline
$^{59}$Ni &  0$^+$        &   8.999       &  4.0          &  13400(900)    &    2536(520)      &     2030(800)                       \\
$^{60}$Ni &  3/2$^-$      &  11.389       &  4.3          &   2000(700)    &    5249(2031)     &     2200(700)                       \\
\hline     
\hline
\end{tabular}
\\
\label{table:level_parameters}
\end{table*}
%----------------------------------------------------------------------------------------

Assuming the spin distribution of Eq.~(\ref{eq:spin_dist}) and parity symmetry at gives the total NLD at $S_n$
\begin{equation}
\rho(S_n) =  \\
\frac{2\sigma^2}{D_0} \frac{1}{(I_t+1) \mathrm{exp}[-(I_t+1)^2/2\sigma^2]+ I_t\mathrm{exp[-I_t^2/2\sigma^2]}}, \\
\label{eq:rho_Sn}
\end{equation}
where $D_0$ is the level spacing of the $s$-wave neutrons and $I_t$ is the ground state spin of the target nucleus in the (n, $\gamma$) reaction. In Tab.\ref{table:level_parameters} the parameters used in the normalization of the NLD and $\gamma$SF are listed.

Note that in Eq.~(\ref{eq:rho_Sn}), it is assumed that there is an equal number of positive and negative parity states around $S_n$. This assumption has to be considered carefully in the case of these two light nuclei, which both display a strong asymmetry of parity at low excitation energies, where $^{60}$Ni has purely positive parity states below $\approx$ 4.5 MeV, while $^{59}$Ni is dominated by negative parity states at low excitation energies.

In Ref.~\cite{Kalmykov_parity}, NLDs of $J^{\pi} = 2^{+}$ and $2^{-}$ states extracted from studies of $E2$ and $M2$ giant resonances in $^{58}$Ni and $^{90}$Zr are used to test predictions of a parity dependence predicted by the Hartree-Fock-Bogoliubov (HFB) model and shell-model Monte Carlo calculations. The authors of~\cite{Kalmykov_parity} observed no parity dependence around $S_n$, experimentally, in contrast to the model predictions for $^{58}$Ni. Recently, parity-dependent level density in $^{58}$Ni has been calculated using a stochastic estimation with the shell model~\cite{Shimizu}. The calculations are in good argreement with the experimental results in~\cite{Kalmykov_parity} and shows an equilibration of $J^{\pi} = 2^{+}$ and $2^{-}$ states at $E_x \geq$ 8 MeV. 
These investigations support our assumption that there is practically no parity asymmetry at $S_n$ in both $^{59,60}$Ni. 

The NLD of a chain of nickel isotopes, $^{59-64}$Ni, have previously been measured in lithium-induced proton evaporation reactions~\cite{AVoinov_evap1, AVoinov_evap2}. These NLDs are not normalized to a calculated or deduced NLD at $S_n$, they are scaled to match discrete levels at low excitation energy. However, the accuracy of the slope of the NLD depends on the uncertainties in the particle transmission coefficients. 
   
In Fig.~\ref{fig:oslo_voinov} the normalized nuclear NLDs from the current experiment are shown together with data from Ref.~\cite{AVoinov_evap1, AVoinov_evap2}. The agreement between the two datasets is quite good. 

%As reported for many nuclei investigated with the Oslo method ~\cite{Magne_levelReview}, the level densities exhibit a constant temperature behaviour. %%%AC sier ikke så veldig constant temperature. WHAT parameters did you use for the interpolation?  

%  \begin{figure}
% \begin{center}
% \includegraphics[clip,width=\columnwidth]{Figures/counting5960.pdf}
%\vskip 2cm
%\caption{(Color online) Comparison of the two isotopes.}
%\label{fig:counting_5960}
%\end{center}
%\end{figure}

%-----------------------------------------STRENGTH FUNCTIONS----------------------------------------------%

\subsection{$\gamma$SFs}
\label{sec:gamma}
The last step in the normalization procedure is to determine a scaling parameter B for the transmission coefficient. The total radiative width $\langle \Gamma_{\gamma} \rangle$ at $S_n$ for initial spin $I$ and parity $\pi$ is given by~\cite{LarsenSyst}:
\begin{align}
\langle \Gamma_{\gamma}(S_n, I_t \pm 1/2, \pi_t \rangle =  \nonumber \\
\frac{D_0}{4\pi} \int_{E_{\gamma}=0} ^{S_n}\textrm{d}E_{\gamma}\mathrm{B}\mathcal T (E_{\gamma}) \rho(S_n-E_{\gamma}) \nonumber \\
\times \sum_{I=-1}^{1} g(S_n-E_{\gamma}, I_t \pm 1/2 +I)
\label{eq:Gg}
\end{align}
where $I_t$ and $\pi_t$ are the spin and parity of the target nucleus in the $(n, \gamma)$ reaction, and $\rho(S_n-E_{\gamma})$ is the experimental NLD. Here we encounter the question of parity asymmetry again. In Eq.(\ref{eq:Gg}) we assume that the level density $\rho(S_n-E_{\gamma})$ has no asymmetry, as we argued was the case for $\rho(S_n)$. However, we know that both Ni isotopes display strong parity asymmeties at lower excitation energies. Although we realize that the parameter B will be quite uncertain due to the possible parity asymmetry, and uncertainties in the listed values of the $\langle \Gamma_{\gamma}\rangle $s we nevertheless use Eq.~(\ref{eq:Gg}) to provide an estimate of B. Assuming that statistical decay is dominated by dipole transitions, as strongly supported by experimental data~\cite{AC_dipole,Kopecky}, the $\gamma$SF, $f(E_\gamma)$, can be calculated from $\mathcal T (E_{\gamma})$ using~\cite{RIPL3}
\begin{equation}
f(E_\gamma) = \frac{\mathcal T (E_{\gamma})}{2\pi E_{\gamma}^3}.
\label{eq:dipole}
\end{equation}  

\begin{figure}[tb]
\begin{center}
\includegraphics[clip, width=\columnwidth]{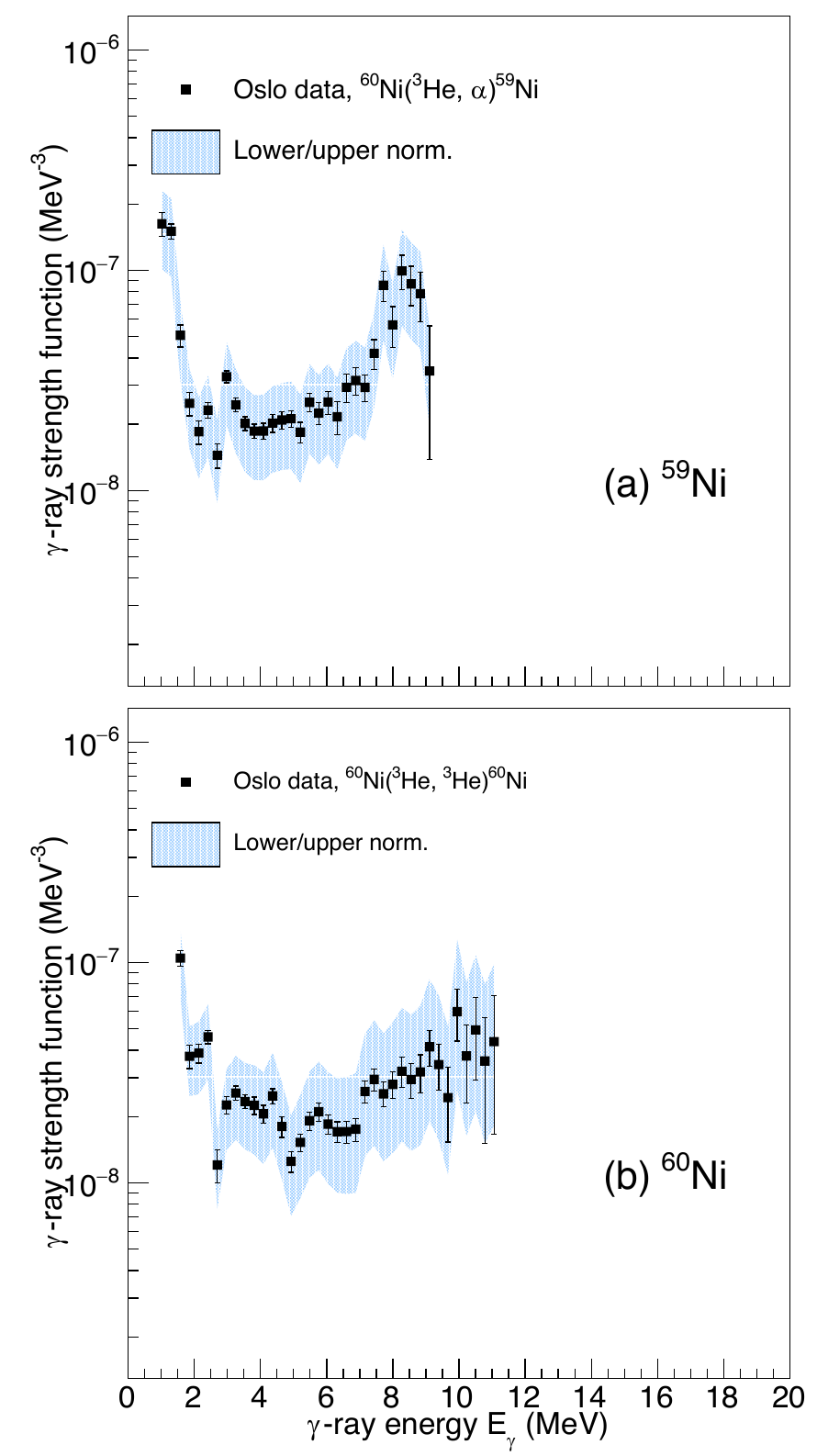}
%\vskip 2cm
\caption{(Color online) $\gamma$SFs of (a) $^{59}$Ni and (b) $^{60}$Ni from the present experiment. The light blue shaded region represent the systematic uncertainties in the analysis, mainly stemming from the uncertainties in the normalization parameters.}
\label{fig:strength}
\end{center}
\end{figure}

The $\gamma$SFs of $^{59, 60}$Ni are presented in Fig.~\ref{fig:strength}, panel (a) and (b), respectively. The light blue colored bands represent the uncertainty band obtained by combining the uncertainties in the listed $\langle \Gamma_{\gamma} \rangle$ values and the uncertainties in the $D_0$ or $D_1$ values. 
The sharp ridges in the $\gamma$SF of $^{59}$Ni at $E_{\gamma}$$\sim$2.5 MeV and at $E_{\gamma}$$\sim$2.5 and 5 MeV in $^{60}$Ni should not be interpreted as physical structures, they are a result of over-subtraction in the first-generation method. In $^{59}$Ni, the points between 8 and 9.5 MeV make up a peak structure; this is likely a result of non-statistical transitions directly to the ground state. The data points in this region will be omitted from here on. In the case of $^{60}$Ni, the data points above $\sim$10 MeV are extracted from a region of the primary $\gamma$-ray matrix that suffers from low statistics. In the rest of this paper, we will exclude these data points.    

In order to strengthen our determination of the absolute value of the $\gamma$SFs, new data is required. For that purpose, we have performed photo-neutron measurements on $^{60}$Ni and $^{61}$Ni. Pre-existing data for $^{60}$Ni display fairly large fluctuations in the area proximate to $S_n$. We therefore chose to re-measure the photo-neutron cross sections in order to ensure accuracy in the areas close to $S_n$. Because $^{59}$Ni is an unstable isotope, our best candidate was the only odd, stable Ni isotope, $^{61}$Ni, for which there exist no previous measurents. Hence, we will compare $\gamma$SF data from $^{59}$Ni below $S_n$ with $\gamma$SF for $^{61}$Ni above $S_n$. 

%--------------------------------------------------------------------------------------------------------------------------------------------------------------------------------------------------------
%--------------------------------------------------------------------------------------------------------------------------------------------------------------------------------------------------------

%----------------------------------------g,n-------------------------------------------------------------------%
\section{photo-neutron measurements}
\label{sec:photoneutron}

\subsection{Experimental procedure} 
\label{sec_exp}
\label{sec:to}
%Referer til en artikkel der det står om den smale gamma spreaden! 
The photo-neutron measurements on $^{60,61}$Ni were performed at the NewSUBARU synchrotron radiation facility as a part of an experimental campaign to measure $^{58, 60, 61, 64}$Ni ($\gamma$, n) cross sections.   

The NewSUBARU facility provides quasi-monochromatic $\gamma$-ray beams produced in laser Compton scattering (LCS) between laser photons and relativistic electrons. A series of $\gamma$-ray beams in the energy range $E_{\gamma}\in [8, 20]$ MeV were provided from collisions between relativistic electrons in the energy range $E_e\in [663, 1061]$ MeV and laser photons with a wavelength of 1064 nm, covering the energy range between $S_n$ and $S_{2n}$ of the two Ni isotopes.
The energy profiles of the produced $\gamma$ ray beams were measured with a $3.5^{\prime\prime}\times 4.0^{\prime\prime}$ LaBr$_3$:Ce (LaBr$_3$) detector. The measured LaBr spectra were reproduced by a GEANT4 based code~\cite{GEANT1, GEANT2}, which takes into account the kinematics of the LCS process, including the beam emittance and the interactions between the LCS beam and the LaBr$_3$ detector. The energy profiles of the incident $\gamma$-ray beams could hence be established. 

The samples, $^{60,61}$Ni, had an areal density of $1036$~mg/cm$^2$ and $609$~mg/cm$^2$, respectively. The corresponding enrichments of the two isotopes were $99.5\%$, and $91.14\%$.

For neutron detection, a high-efficiency $4\pi$ detector was used, consisting of 20 $^3\rm{He}$ proportional counters, arranged in 3 concentric rings and embedded in a 36 $\times$ 36 $\times$ 50 cm$^3$ polyethylene neutron moderator. To ensure supression of background neutrons, the surface of the moderator was covered with cadmium lined plates of polyethylene.

The LCS $\gamma$-ray flux was monitored by a $8^{\prime\prime}\times 12^{\prime\prime}$ NaI:Tl (NaI) detector during neutron measurement runs. The number of incoming $\gamma$ rays per measurement was estimated using the “pile-up” technique~\cite{Kondo, Dan}. 
%See Fig.\ref{fig:setup} for a schematic illustration of the experimental set up.

The total photo-neutron cross section, $\sigma_{\rm{exp}}$, for an incoming beam with maximum $\gamma$-energy $E_{Max}$ is given by,
\begin{equation}
\sigma_{\rm{exp}}=\int_{S_n}^{E_{\rm{Max}}}n_{\gamma}(E_{\gamma})\sigma(E_{\gamma})dE_{\gamma}=\frac{N_n}{N_tN_{\gamma}\xi\epsilon_n g},
\label{eq:cross1}
\end{equation}
where $n_{\gamma}(E_{\gamma})$ gives the normalized energy distribution of
the $\gamma$-ray beam simulated as described above and $\sigma(E_{\gamma})$ is the actual photo-neutron cross section. Further, $N_n$ represents the number of
neutrons detected, $N_t$ gives the number of target nuclei per unit area, $N_{\gamma}$
is the number of $\gamma$ rays incident on target, $\epsilon_n$ represents the
neutron detection efficiency, and $\xi=(1-e^{-\mu t})/(\mu t)$ gives a
correction factor for self-attenuation in the thick target measurement. Here, $\mu$ represents the mass
attenuation coefficient (in cm$^2$/g), tabulated in Ref.~\cite{NIST}. The factor $g$ represents the fraction of the $\gamma$ flux above with energy above neutron emission threshold.

Throughout the experiment, the laser was on for 80 ms and off for 20 ms in every 100 ms, in order to measure background neutrons and $\gamma$-rays.

The total uncertainty in the measurements consists of $3.2\%$ from neutron detection efficiency, $3\%$ in the number of $\gamma$-rays, and the statistical uncertainty in the number of measured neutrons.

\subsection{Data analysis}  
We need to determine the photo-neutron cross section as a function of $E_{\gamma}$, $\sigma(E_{\gamma})$, included in the integral of Eq.(\ref{eq:cross1}). Each of the measurements correspond to a specific integral, relating $\sigma(E_{\gamma})$ to the measured beam profile and the experimentally measured cross section. By discretizing the equations, we get
\begin{equation}
\bf{\sigma}_{\rm{folded}}=\bf{n_{\gamma}}\bf{\sigma}
\end{equation}
or, more explicitly,
\begin{equation}
\begin{pmatrix}\sigma_{\rm{folded1}}\\\sigma_{\rm{folded2}}\\ \vdots \\ \sigma_{\rm{foldedN}} \end{pmatrix}\\\mbox{}=
\begin{pmatrix}n_{\gamma 11} & n_{\gamma 12}& \cdots &\cdots &n_{\gamma 1M} \\ n_{\gamma 21} & n_{\gamma 22}& \cdots & \cdots &n_{\gamma 1M} \\ \vdots &\vdots & \vdots & \vdots & \vdots \\ n_{\gamma N1} & n_{\gamma N2}& \cdots & \cdots &n_{\gamma NM}\end{pmatrix}
\begin{pmatrix}\sigma_{1}\\\sigma_{2}\\ \vdots \\ \vdots \\\sigma_{M} \end{pmatrix}
\label{eq:matrise_unfolding}
\end{equation}

Here, $\bf{\sigma_{folded}}$ and $\bf{\sigma}$ represent the iterate and the actual cross section, respectively. Each row in $\bf{n_{\gamma}}$ corresponds to a GEANT 4 simulated $\gamma$ beam profile belonging to a specific energy of the electron beam. The system of linear equations in Eq.(\ref{eq:matrise_unfolding}) is underdetermined, so the true $\sigma$ cannot be found using matrix inversion. In order to find $\sigma$,
we use a folding iteration method. The main features of this method are as follows:
%To find the most probable solution, we utilize a folding iteration method. 

\begin{itemize}

\item [1)] As a starting point for the iterations, we choose $\bf{\sigma}$ = $\bf{\sigma_{0}} = C$, where $C$ is a constant. This vector is multiplied with $\bf{n_{\gamma}}$ and the first iterate is produced, $\bf{\sigma_{\rm{folded0}}}$.
\item[2)] In order to establish the next input function, $\bf{\sigma_1}$, we add the difference of the experimentally measured spectrum, $\bf{\sigma_{\rm{exp}}}$, and the folded spectrum,$\bf{\sigma_{\rm{folded0}}}$, to $\bf{\sigma_0}$. This way of choosing the subsequent input functions is known as the difference approach. In order to be able to add the folded and the input vector together, we first perform a spline
fit on the folded vector, then interpolate so that the two vectors have equal dimensions. Our new input vector is:

\begin{equation}
\sigma_1 = \sigma_0 + (\sigma_{\rm{exp}}-\sigma_{\rm{folded0}}).
\end{equation} 

\item[3)] The steps 1) and 2) are repeated until convergence is achieved, that is $\sigma_{\rm{folded}}\approx \sigma_0$. 
\end{itemize}

\begin{figure}[tb]
\begin{center}
\includegraphics[clip, width=\columnwidth]{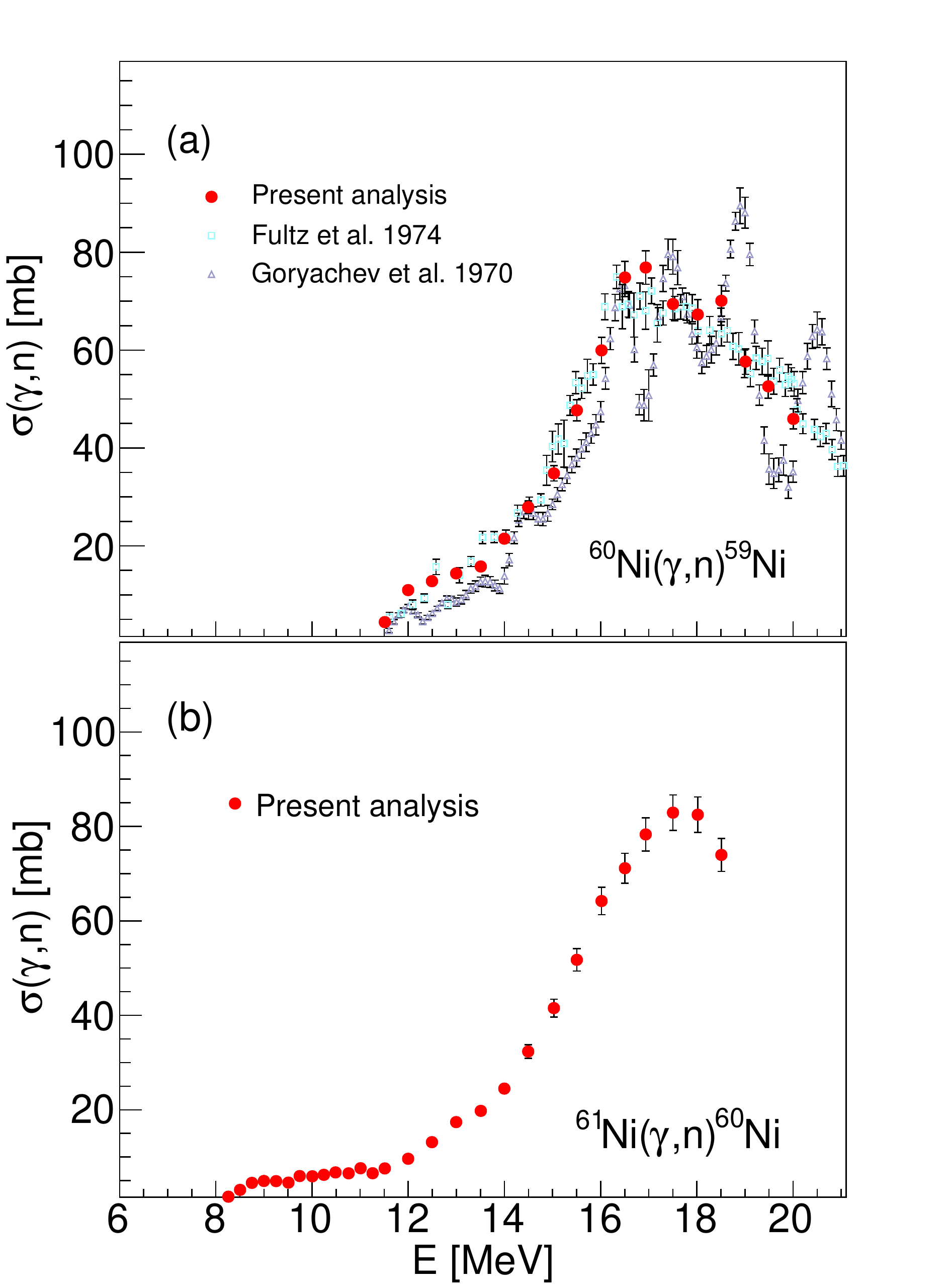}
%\vskip 2cm
\caption{(Color online) A comparison between existing photo-neutron cross sections and the present data on $^{60}$Ni in (a) and photo-neutron cross section of $^{61}$Ni in (b).}
\label{fig:cross_sec}
\end{center}
\end{figure}

The statistical errors in the measurements of neutrons are $\sim$ 1$\%$, except for data points very close to $S_n$, where they can reach 3-4$\%$. These statistical fluctuations in the measured cross sections,$\sigma_{\rm{exp}}$, can not easily be distinguished from actual, small structures in the measurements. To avoid introducing strong fluctuations in the unfolded data caused by statistical fluctuations, we perform a smoothing of the unfolded cross sections in each iteration. The energy-dependent smoothing widths used in the analysis of the Ni isotopes are between 200 and 400 keV, corresponding to the approximate FWHM of the incoming $\gamma$-ray beams. After $\approx$ 5 iterations, $\sigma_{\rm{folded}}$ $\approx$ $\sigma_0$. The systematic uncertainty in the measured cross sections mainly come from the uncertainty in the efficiency calibration of the neutron detector and the estimation of the $\gamma$-ray flux. In order to account for this uncertainty, we estimate an upper and lower limit of the measured cross sections, and unfold these limits. 
In Fig.\ref{fig:cross_sec} (a) and (b) the resulting unfolded ($\gamma$, n) cross sections of $^{60,61}$Ni are presented. Here, the unfolded cross sections are evaluated at $E_{\max}$ of the individual $\gamma$-ray beams. The error bars represent the systematic and statistical uncertainties combined. For $^{60}$Ni, our cross sections are compared with existing cross sections from Refs.~\cite{Fultz, Goriachev}. The present cross sections are on average higher than those reported by Goryachev et al.~\cite{Goriachev}, except in regions where the latter show large fluctuations. The agreement with the data measured by Fultz et al.~\cite{Fultz} is rather good.

\section{Comparing $\gamma$SF below and above $S_n$}

The $\gamma$SF deduced from the Oslo data is complemented by the ones extracted from the photoneutron cross section through the expression 
\begin{equation}
f(E_{\gamma}) = \frac{1}{3\pi^2\hbar^2c^2}\frac{\sigma(E_{\gamma})}{E_{\gamma}}. 
\end{equation}

\begin{figure}[tb]
\begin{center}
\includegraphics[clip, width=\columnwidth]{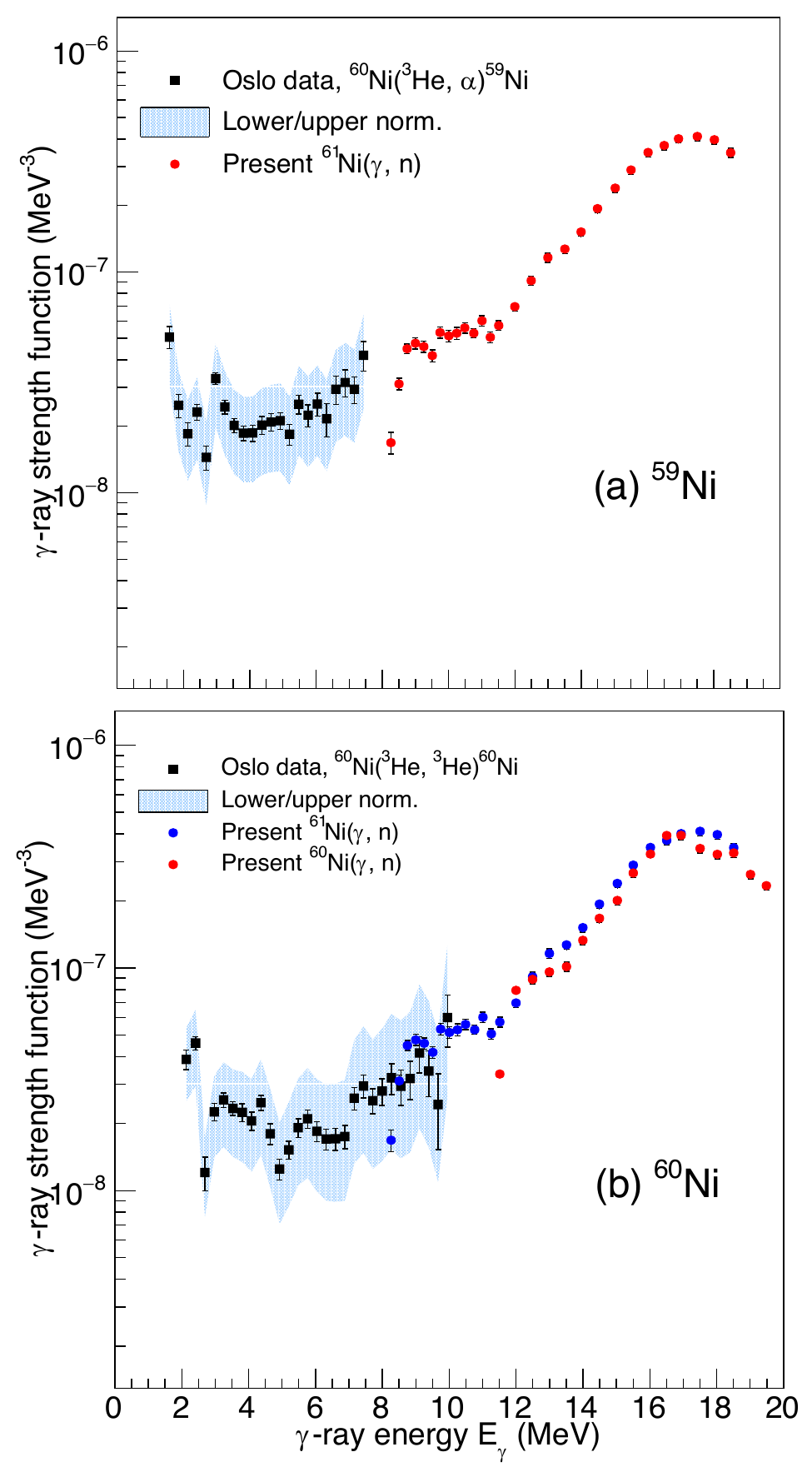}
%\vskip 2cm
\caption{(Color online) Combining the $\gamma$SF above and below $S_n$ for $^{60}$Ni in (a). In (b) both $\gamma$SFs from $^{61}$Ni and $^{60}$Ni above $S_n$ are combined with the $\gamma$SF of $^{59}$Ni for the region below $S_n$.}
\label{fig:photo_oslo}
\end{center}
\end{figure}

%The newly deduced cross sections of $^{60}$Ni is shown together with previously measured data. Previously f(E1) and f(M1) have been measured for $^{61}$Ni at $E_{\gamma}$ = 7.8 MeV, see Fig.\ref{fig:photo_oslo}.
In Fig.~\ref{fig:photo_oslo} (a) and (b) we combine the photoneutron data above $S_n$ with the data from the Oslo Method. The photoneutron data exhibit extra strength in the region between 9 and 12 MeV. Small fluctuations in the $\gamma$SF is also seen in this region. Djajali et al.~\cite{Djalali_spinflip} show a highly fragmented spin-flip resonance for $^{60,62}$Ni between $\approx$ 8 and 11 MeV in the (p,p$^{\prime}$) reaction, and average resonance capture data from Ref.~\cite{RIPL2} show a dominance of $M1$ strength compared to the $E1$. In light of this, the extra strength in $^{61}$Ni could be interpreted as a strong, fragmented spin-flip resonance. The photoneutron data on $^{60}$Ni show a splitting around the peak of the GDR and also an indication of extra strength a few MeV above $S_n$. The Oslo data exhibit extra strength between $\sim$7.5 and 10 MeV. This could be due to enhancement of $E1$ strength as reported in~\cite{Scheck_rapid, Scheck_full} or/and an $M1$ spin-flip.

Both Ni isotopes exhibit a low-energy enhancement in the $\gamma$-ray strength for $\gamma$ energies below $\sim$4 MeV. 

\section{Theoretical considerations}
We are not able to decompose the experimental $\gamma$SFs extracted from the photo-neutron or charged particle reactions in an $M1$ and $E1$ part. To this end, we rely on theoretical approaches. In the last part of the paper we present shell model and QTBA calculations of the $M1$ and $E1$ part, respectively, of the $\gamma$SFs of $^{59,60}$Ni.  

%-----------------------------------------SHELL MODEL CALCULATIONS----------------------------------------------%
\subsection{Shell model calculations of the M1 strength}
\label{sec:shellmodel}
The original definition of the $\gamma$SF~\cite{Barto}, can be reformulated for M1 transitions as follows
\begin{equation}
f_{M1}(E_{\gamma}, E_{i}, I_i, \pi_i) = a \left < B(M1)\right > (E_{\gamma}, E_{i}, I_i, \pi_i) \rho(E_{i}, I_i, \pi_i))
\end{equation}
where $a = 11.5473 \times 10^{-9} \mu_{N}^{-2}$ MeV$^{-2}$, and $\left < B(M1)\right >$ and $\rho(E_{i}, I_i, \pi_i))$ are the average reduced transition strength and partial level density, respectively, of states with a given excitation energy $E_i$, spin $I$, and parity $\pi$. Shell model calculations have been used to calculate excited states in $^{59,60}$Ni and $B(M1)$ strengths in the relevant energy region. The shell model codes {\scshape KSHELL}~\cite{KSHELL} and NUSHELLX@MSU~\cite{NuShell} have been utilized. For the KSHELL calculations, we used the CA48MH1 interaction, which contains the orbitals $\pi(p_{3/2}p_{1/2}f_{5/2}f_{7/2})$ and $\nu(p_{3/2}p_{1/2}f_{5/2}g_{9/2})$. For calculations with this model space, we restrict the maximum number of excited protons from the $f_{7/2}$ orbital to 2, but use no truncation on the neutrons. We calculate all accessible states with $I \in \left[0, 14\right]$ and $I \in \left[1/2, 29/2\right]$, in the case of $^{60}$Ni and $^{59}$Ni, respectively, of both parities.   

For the NuShellX calculations, we used the {\scshape GPFX1A} interaction~\cite{hamilton_1, hamilton_2} for the $pf$ shell. The model space for $^{59, 60}$Ni was $(0 f_{7/2})^{8-t_p}(0 f_{5/2}, 1 p_{3/2}, 1 p_{1/2})^{t_p}$ for protons, where $t_p$ = 0$\textendash$3 and $(0 f_{7/2})^{8-t_n}(0 f_{5/2}, 1 p_{3/2}, 1 p_{1/2})^{n+t_n}$ for neutrons, where $t_n$ = 0$\textendash$2 and n = 3 and n = 4 for $^{59}$Ni and $^{60}$Ni, respectively. For this interaction, we calculate all accessible states with $I \in \left[0, 14\right]$ and $I \in \left[1/2, 29/2\right])$, in the case of $^{60}$Ni and $^{59}$Ni, respectively. In the case of $^{60}$Ni and $^{59}$Ni only positive and negative parity states, respectively, are accessible.

\begin{figure}[tb]
\begin{center}
\includegraphics[clip, width=\columnwidth]{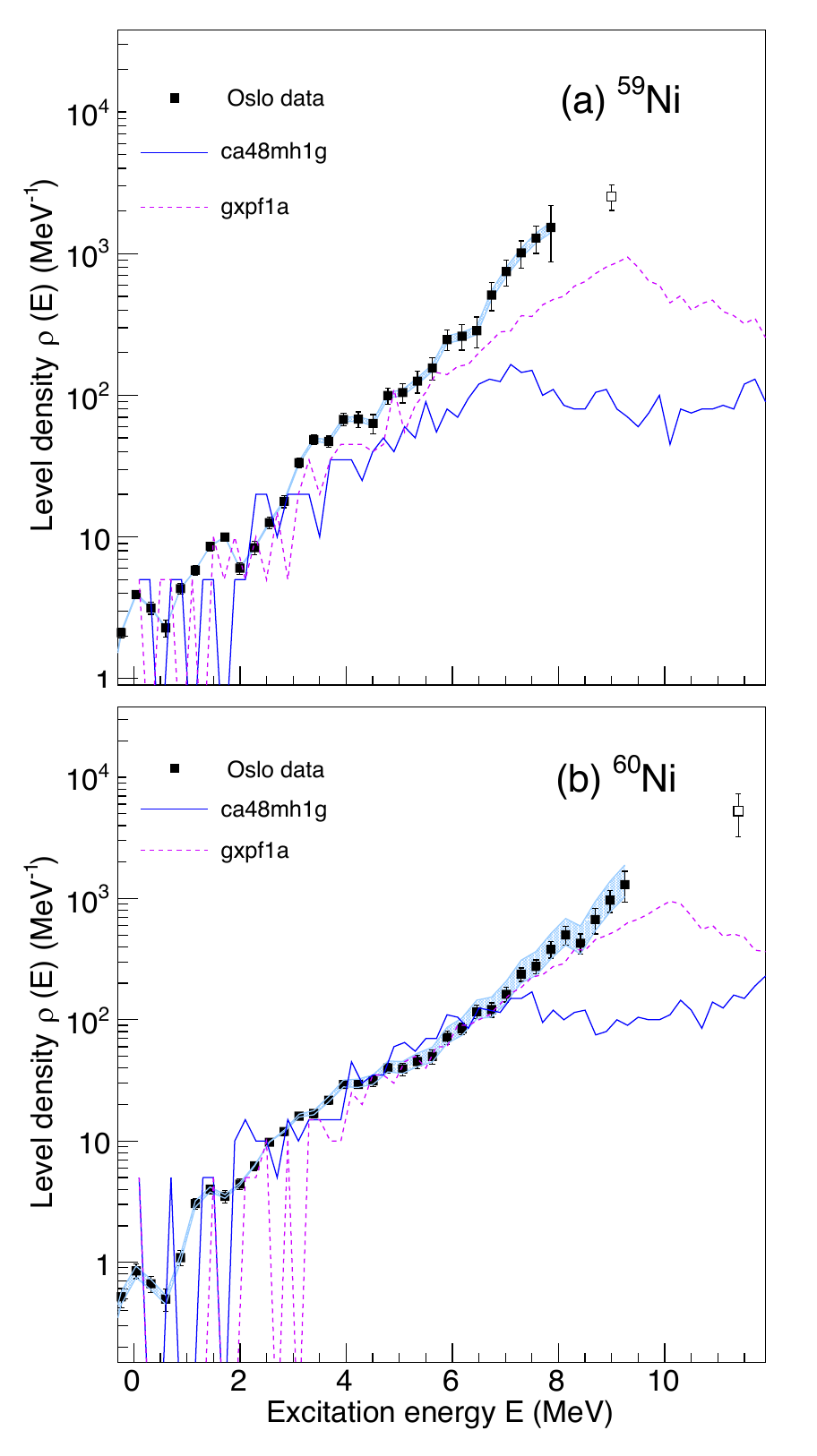}
%\vskip 2cm
\caption{(Color online) Shell-model level densities for (a) $^{59}$Ni (b) $^{60}$Ni, calculated using the ca48mh1g interaction (blue, solid line) and the  gxpf1a interaction (dotted, purple line) compared to experimental data (black filled squares) from the current experiment.}
\label{fig:Shell_level}
\end{center}
\end{figure}
%-----------------------------------------------------------------------------------------

The obtained level densities, summed over all spins and accessible parities, are shown in Fig.~\ref{fig:Shell_level} and compared to experimental data. For $^{60}$Ni there is a very good agreement between the GXPF1A calculations and measured level density up to E$\approx$8.5 MeV. For higher excitation energies, the calculations are consistently smaller than the measured one. We interpret this as a direct effect of the fact that only positive parity states have been calculated, and that in this excitation energy range, the number of negative parity states are expected to increase rapidly, we thus start to see an effect of leaving out the negative parity when calculating the total level density. Around 10 MeV, a different effect is seen as a kinck in the plotted GXPF1A calculations. Here we see a direct effect of the fact that we chose to calculate the first 300 states of each spin, and some spins are exhausted. The general trends are the same in  $^{59}$Ni, but the discrepancy between the calculations and the measured values is overall larger. This is due to the fact that the truncations in the model space have a more pronounced effect due to the unpaired neutron in $^{59}$Ni. 

According to the generalized Brink hypotesis, $f_{M1}(E_{\gamma}, E_{i}, I_i, \pi_i) \approx f_{M1}(E_{\gamma})$. Thus we obtain $f_{M1}(E_{\gamma})$ by averaging over $E_i$, $I$ and $\pi$. Note that we only include $E_i$, $I$ and $\pi$ bins where $f_{M1}$ is non-zero in the average.   

\begin{figure}[tb]
\begin{center}  
\includegraphics[clip, width=\columnwidth]{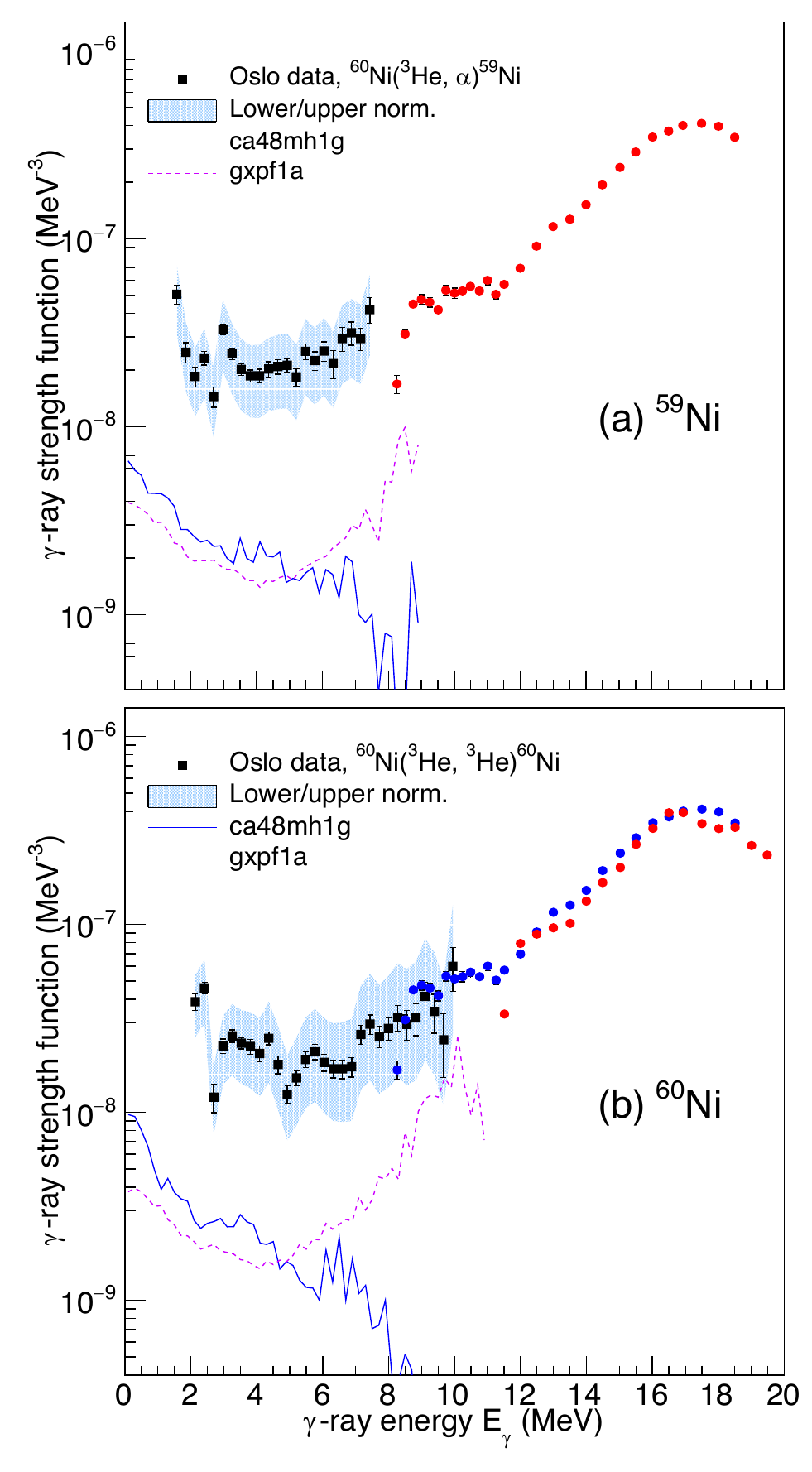}
%\vskip 2cm
\caption{(Color online) Shell-model $M1$ $\gamma$SFs for (a) $^{59}$Ni (b) $^{60}$Ni, calculated using the ca48mh1g interaction (blue, solid line) and the gxpf1a interaction (dotted, purple line) compared to experimental data (black filled squares) from the current experiment.}
\label{fig:Shell_strength}
\end{center}
\end{figure}

The extracted strength functions, $f_{M1}(E_{\gamma})$, are shown in Fig.~\ref{fig:Shell_strength}. Both calculations show an enhancement peaking at $E_{\gamma}$ = 0 MeV, in accordance with the shape the experimental data. However, the absolute values of the two calculations differ. At $E_{\gamma}\approx$ 0 MeV, the CA48MH1G calculations have a value roughly twice as large as the GXPF1A calculations for $^{60}$Ni. The B(M1) values were in both cases calculated using effective $g_s$ factors of $g_s^\textrm{eff} = 0.9 g_s^\textrm{free}$. However, since the CA48MH1G model space is not $LS$ closed, {\it i.e.}\ it is truncated across spin-orbit partners, it may be necessary to apply a higher quenching of the $g_s$ values~\cite{Homna}. In the energy region $E_{\gamma} \geq$ 5 MeV the the GXPF1A result exhibits a large structure. This could possibly stem from transitions between the spin-orbit partners $f_{7/2}$ and $f_{5/2}$. As the $\nu f_{7/2}$ orbital is not present in the model space of the CA48MH1G calculations this could explain the difference.
% Here, it becomes apparent that the $M1$ strength is rather sensitive to the choice of interaction and truncations in the model space.  

From Fig.\ref{fig:Shell_strength}, we see that the calculated $M1$ strength is not sufficient to describe the total $\gamma$SF in the low-energy range. In a recent publication by Jones et al.~\cite{Jones}, results suggest a mixture of $M1$ and $E1$ radiation in the enhancement region, with a small magnetic bias between 1.5 and 2 MeV. Shell-model calculations of the $E1$ component of the $\gamma$SF reported by K. Sieja~\cite{KSieja} show a quite strong constant $E1$ component at low $\gamma$ energies in the case of $^{44}$Sc. If this trend in the $E1$ radiation is also present in the Ni isotopes, it would account for some of the discrepancy.  
 %-----------------------------------------------------------------------------------------
%-----------------------------------------QTBA CALCULATIONS----------------------------------------------%

\subsection{QTBA calculations of the $E1$ strength}
\label{sec:QTBA}

The Quasiparticle Time Blocking Approximation (QTBA) is formulated in~\cite{Tsel}, therefore we will in the following only summarize the crucial equations. 
The conventional 1p1h Random Phase Approximation (RPA) written in the configuration space of the single-particle
wave functions $\phi_{\nu}$ has the form:

\begin{equation}
(\epsilon_{\nu_1}-\epsilon_{\nu_2}-\Omega)\chi^m_{\nu_1\nu_2}=(n_{\nu_1}-n_{\nu_2})\sum_{\nu_3\nu_4} F^{ph}_{\nu_1\nu_4\nu_2\nu_3}\chi^m_{\nu_3\nu_4}.
\label{eq:1p1h}
\end{equation}

Here, $\epsilon_{\nu}$ are the single-particle energies, $F^{ph}$ is the residual ph interaction and, $n_{\nu}$ are the occupation numbers. In the self-consistent approach
all these input data follow from the mean-field solution.
From Eq.(~\ref{eq:1p1h}) one obtains the excitation energy $\Omega$ of an even-even nucleus and the corresponding ph-transition matrix elements $\chi^m_{\nu_1\nu_2}$.

For numerical applications, it is more convenient to solve the equation in \textit{r}-space because this allows for a more efficient treatment of the continuum. Instead
of the homogeneous integral in Eq.(~\ref{eq:1p1h}) one solves an inhomogeneous equation of the form:

\begin{equation}
\begin{split}
\rho(r,\Omega)=-\int d^3r'A(r,r',\Omega)Q^{eff}(r',\Omega)\\
-\int  d^3r' d^3r''A(r,r',\Omega)F^{ph}(r',r'')\rho(r'',\Omega),
\end{split}
\label{eq:inhomo}
\end{equation}
where $Q^{eff}(r,\Omega)$ is an external field and $A(r,r',\Omega)$ is
the ph-propagator in the \textit{r}-space. The poles of this equation
are the excitation energies of an even-even nucleus,
and $\rho(r,\Omega)$ at a given pole is the corresponding transition
density.

\begin{figure}[t]
\includegraphics[width=7cm,clip]{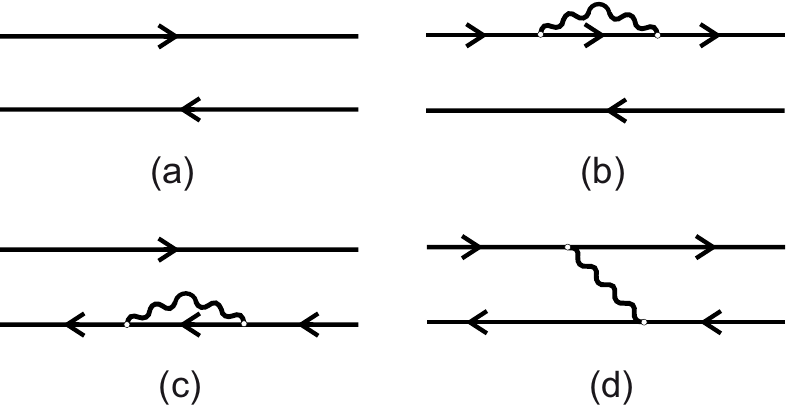}
\caption{An illustration of the QTBA idea. Graph (a) denotes the ph-propagator of the RPA. Corrections due to phonons are indicated in (b-d).
The graphs (b) and (c) are corrections to the propagator and (d) is a contribution to the ph-interaction. The wavy lines and
the solid lines denote the phonons and the single-particle propagators, respectively. See text for details.}
\label{fig-1}       % Give a unique label
\end{figure}

The inclusion of phonons gives rise to a modification of
the single-particle energies, see Fig.(~\ref{fig-1}), graphs (b,c). It also gives rise to a modification
of the ph-interaction, which is shown in Fig.~\ref{fig-1}, graph (d). The modification of the propagator as shown in Fig.~\ref{fig-1} is the essence of the QTBA approach, see~\cite{Tsel, PhysRep}. A further advantage of the \textit{r}-space is that the structure of
Eq.(~\ref{eq:inhomo}) is not changed if phonons are included, only the propagator $A$ is modified, and phonons and the ph interaction $F^{ph}$ are accounted for self-consistently.

Although the QTBA model~\cite{Tsel} was originally formulated in terms of this general basis, several simplifications are performed in our calculations.
Namely, the QTBA approach is designed to use the Bardeen, Cooper and Schrieffer (BCS)-based quasiparticle basis and we use the Hartree-Fock-Bogoliubov (HFB) approach to extract the quasiparticle characteristics and corresponding
wave functions (i.e., the occupation numbers are treated as for the BCS approximation). The spin-orbit residual interaction is
dropped. The velocity-dependent terms of the Skyrme force are approximated by their Landau-Migdal limit, although
some more physically sound modifications are included. There are two kinds of velocity-dependent terms: the first one is
 $\propto \textbf{k}^2 \delta(\textbf{r}-\textbf{r}')$ and the second one is 
 $\propto \textbf{k}^{\dagger} \delta(\textbf{r}-\textbf{r}')\textbf{k}$ (P-wave interaction in momentum space). The averaged value over
the density of the first term gives $k^2_F/2\delta(r-r')$ while that of the second one is zero. Such an approximation violates the
self-consistency and one has to modify the parameters of the residual interaction to obtain the spurious center-of-mass state to
zero. For this reason we only replace the term which is proportional to $t_1 \textbf{k}^2_F\delta(\textbf{r}-\textbf{r}')$ up to 25\% as we approximate this term.

In general, the QTBA completely accounts for the single-particle continuum at the RPA level for magic nuclei and includes the effect of ground-state
correlations caused by phonon couplings (PC)~\cite{PhysRep}. However, because of technical difficulties connected with pairing, these effects are not considered in the present calculations. We discretized the continuum with quasiparticle
energy cutoff of 100 MeV. We checked that, within this approach, the energy-weighted sum rule (EWSR) is fully exhausted (for the case without the velocity-dependent terms) and that the use of
a larger basis did not bring any noticeable differences. The QTBA calculations are performed with the same basis. 
We use 14--16 low-lying phonons of $L = 2-6$ multipolarity and normal parity. They are obtained within the (Q)RPA with the calculated effective interaction using the same quasiparticle energy
cutoff. Such a consistent method to calculate phonons is the reason why we use a larger number of phonons than in the phenomenological Exteded Theory of finite Fermi systems (ETFFS)~\cite{PhysRep}. 

The ground states are calculated within the HFB approach using the spherical code HFBRAD~\cite{HFBRAD}. To date there are many different Skyrme parameterizations
serving slightly different aims and fitting some bulk properties of the ground state. Here we use the SLy4 parameterization of the Skyrme force~\cite{Chabanat}, which proves to be rather successful
in describing bulk properties of the ground state and some excited states within the (Q)RPA~\cite{Terasaki}. The residual interaction for the (Q)RPA and QTBA calculations is derived as the second derivative of
the Skyrme functional~\cite{Terasaki}. 
%--------------------------------------------------------------------------------------------------------------------
\begin{figure}[tb]
\begin{center}
\includegraphics[clip, width=\columnwidth]{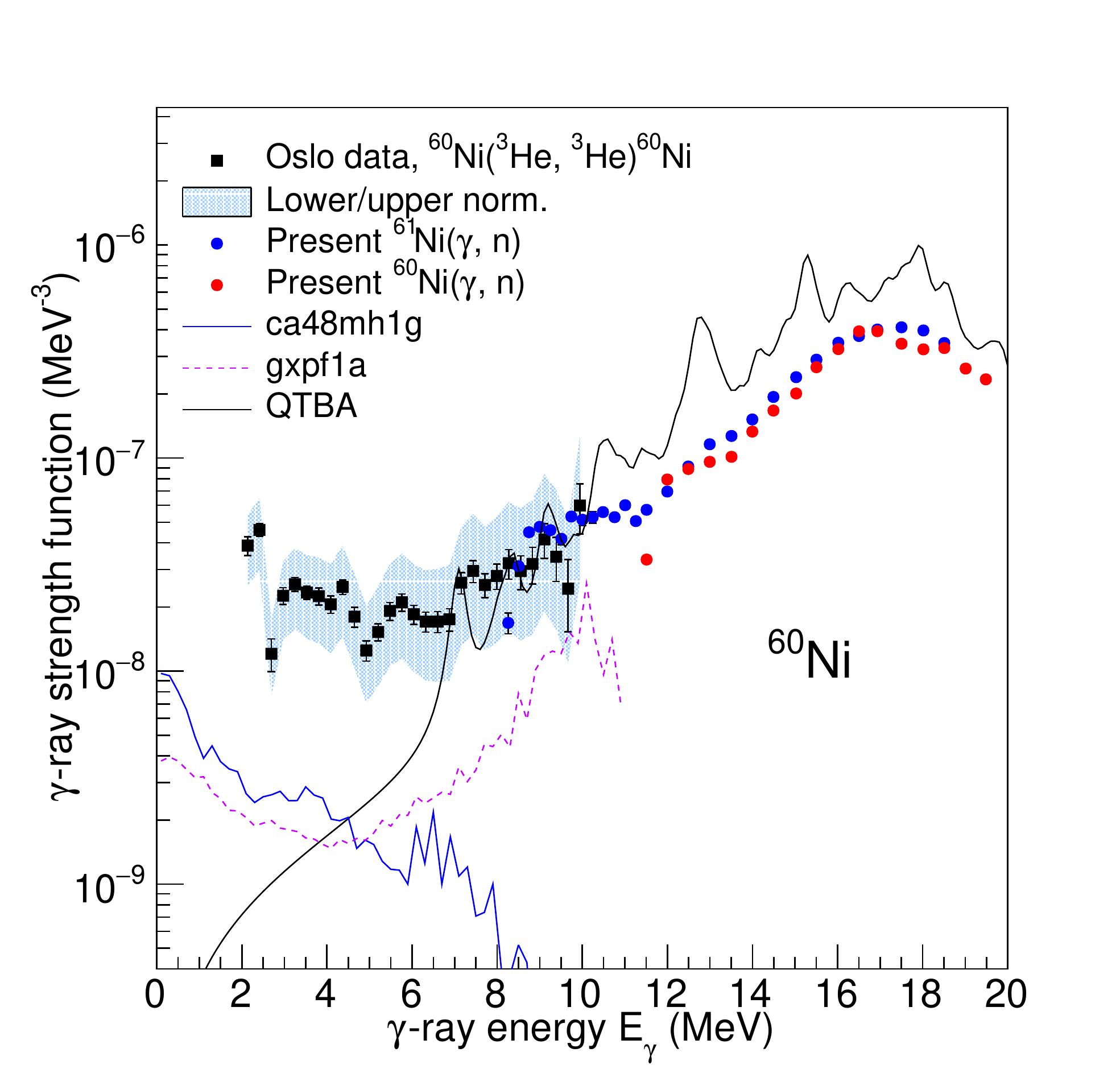}
%\vskip 2cm
\caption{(Color online) QTBA calculations of the $E1$ $\gamma$SF (black, full line) compared with experimental results (same as Fig.~\ref{fig:Shell_strength}.}   
\label{fig:QTBA}
\end{center}
\end{figure}
%--------------------------------------------------------------------------------------------------------------------
The QTBA approach has been very successful in reproducing the excess strength seen in Sn isotopes measured with the Oslo method~\cite{E1_micro}. In the calculations a smoothing parameter of 200 keV is applied, which approximately coincides with the experimental resolution of the Oslo experiments. 

In Fig.~\ref{fig:QTBA} we show the QTBA calculations together with experimental data. The QTBA calculations are in good agreement with the our measured data between 7 MeV and 12 MeV. For higher values of $E_{\gamma}$ the QTBA calculations are consistently higher than the photoneutron data for $^{61,60}$Ni. A possible expalantion for this dicrepancy could be that our experimental data above $S_n$ do not include the ($\gamma$,p) channel, which could be substantial for these light nuclei. 

\section{Conclusions}
\label{sec:conc}
In summary, we have used two different experimental methods to measure $\gamma$SFs of Ni isotopes above and below the neutron separation energy. The $\gamma$SFs of $^{59,60}$Ni extracted using the Oslo method are in good argreement with the strengths of $^{60,61}$Ni measured in photo-neutron experiments. The $\gamma$SFs of $^{59,60}$Ni isotopes display an enhancement at energies below $\sim$ 4 MeV. The $\gamma$SFs of $^{60,61}$Ni above $S_n$ both display extra strength a few MeV above $S_n$. The electromagnetic characters of these structures are experimentally inaccessible in our case, therefore we have compared our data with theoretical calculations of the $\gamma$SF. For the $M1$ part we have performed shell-model calculations. They show an enhancement in the $\gamma$SF for low $E_\gamma$. In absolute strength the shell-model$M1$ strength is considerably lower than the experimental values. This could indicate an extra $E1$ component in the $\gamma$SF. For the $E1$ part, we have performed QTBA calculations. They describe our experimental data well for $E_\gamma$ values between 7 and 12 MeV, but exceed our measured data for higher values of $E_{\gamma}$. Future shell-model calculations will focus on the $E1$ component of the low energy $\gamma$SF.

%The NLDs and $\gamma$SFs of $^{59,60}$Ni have been extracted from particle-$\gamma$ coincidence data using the Oslo method.  The $\gamma$SFs for both Ni isotopes display an enhancement at energies below $\sim$ 4 MeV. The nature of this extra strength could not be determined from the experimental data in this work. However, shell-model calculations of $^{59, 60}$Ni cl$M1$ strength for low $\gamma$-ray energies. 

%\vskip 3 cm
\noindent{\bf Acknowledgements} \\
We would like to thank E.~A.~Olsen, A.~Semchenkov, and J.~Wikne for providing excellent experimental conditions. We also recognize the valuable work of A.~B\"urger in connection with development of the data acquisition and sorting codes at OCL and we would like to thank I.~E.~ Ruud for taking shifts at OCL. A.~C.~Larsen acknowledges funding from ERC-STG-2014, grant agreement no. 637686. G.~M.~Tveten acknowledges funding from the Research Council of Norway, project grant no. 262952. The work of O. Achakovskiy and S. Kamerdzhiev has been supported by the grant of Russian Science Foundation (the project No 16-12-10155). D.~Symochko acknowledge the support of Deutsche Forschungsgemeinschaft through grant No. SFB 1245. A.~Voinov acknowledges support through grant No. DE-NA0002905.

\end{document}